\documentclass[onecolumn]{aa}
\usepackage[varg]{txfonts}
\usepackage{amsmath}
\usepackage{amsfonts}
\usepackage{amssymb}
\usepackage{multirow}
\usepackage{graphicx}
\usepackage{tabularx}
\usepackage[breaklinks=true]{hyperref}
\usepackage{url}

\usepackage{natbib,twoopt}
\bibpunct{(}{)}{;}{a}{}{,}

\def \apjs {{\rm {ApJS}}}

\def \aap {{\rm {A\&A}}}

\def \apjl{\rm {ApJL}}

\begin{document}
\title{Evidence for at least three planet candidates orbiting HD20794}
\author{F. Feng \inst{1}\thanks{\email{fengfabo@gmail.com or f.feng@herts.ac.uk}}\and M. Tuomi\inst{1} \and H. R. A. Jones\inst{1}}
\offprints{Fabo Feng, \email{f.feng@herts.ac.uk, fengfabo@gmail.com}}

\institute{Centre for Astrophysics Research, School of Physics, Astronomy and Mathematics, University of Hertfordshire, College Lane, Hatfield AL10 9AB, UK}

\date{\today}

\abstract{}{We explore the feasibility of detecting Earth analogs around Sun-like stars using the radial velocity method by investigating one of the largest radial velocities datasets for the one of the most stable radial-velocity stars HD20794.}{ We proceed by disentangling the Keplerian signals from correlated noise and activity-induced variability. We diagnose the noise using the differences between radial velocities measured at different wavelength ranges, so-called ``differential radial velocities''.} {We apply this method to the radial velocities measured by HARPS, and identify four signals at 18, 89, 147 and 330\,d. The two signals at periods of 18 and 89\,d are previously reported and are better quantified in this work. The signal at a period of about 147\,d is reported for the first time, and corresponds to a super-Earth with a minimum mass of 4.59 Earth mass located 0.51\,AU from HD20794. We also find a significant signal at a period of about 330\,d corresponding to a super-Earth or Neptune in the habitable zone. Since this signal is close to the annual sampling period and significant periodogram power in some noise proxies are found close to this signal, further observations and analyses are required to confirm it. The analyses of the eccentricity and consistency of signals provide weak evidence for the existence of the previously reported 43\,d signal and a new signal at a period of about 11.9\,d with a semi amplitude of 0.4\,m/s.} {We find that the detection of a number of signals with radial velocity variations around 0.5\,m/s likely caused by low mass planet candidates demonstrates the important role of noise modeling in searching for Earth analogs.}
\maketitle 
\section{Introduction}     \label{sec:introduction}
High precision spectrometers enable us to find super-Earths by measuring the Doppler shift of the stellar spectra caused by the periodic perturbations of exoplanets orbiting the target stars. Unlike the transit method which relies on a particular orbital orientation, the radial velocity (RV) method can provide useful information about planets around all stars. The recent detection of the nearest ``habitable-zone'' planet candidate Proxima Centauri b demonstrates the important role of the RV technique in detecting Earth analogs \citep{anglada16}.

However, the stellar spectrum is contaminated by various noise sources such as stellar activity, instrumental noise and non-perfect data reduction. An analysis of high-cadence spectroscopy of M dwarfs shows that the noise floor (about 1\,m/s) of current Doppler surveys has to a large extent an instrumental origin \citep{berdinas16}. Our analysis is also concerned with demonstration of the strong dependence of activity-induced noise on wavelength. The minimization of both instrumental and stellar noise requires a noise model accounting for noise correlated in time and wavelength. 

Considering that complex models typically lead to false negatives while simple models lead to false positives \citep{feng16}, we follow the Goldilocks principle introduced in \cite{feng16} to select the best noise model for a given RV data set. In other words, we devise a noise model framework to diagnose the noise in a given data set. In this work, we apply this noise model framework to the HARPS RV measurements of HD20794. 

HD20794 is a high-velocity and metal-deficient G8 star with a mass of $0.813^{+0.018}_{-0.012}$\,$M_\odot$ \citep{ramirez13}. It has been reported to host at least 3 planets and a dust disk \citep{pepe11, kennedy15}. The postulated planets are likely super-Earths with orbital periods of 18.1, 40.1 and 90.3\,d \citep{pepe11} (hereafter P11). These results of P11 were obtained using periodogram analysis which assumes white RV noise and circular planetary orbits. Although HD20794 is known to have planets, it has long been chosen as a special RV reference target due to its notable stability over long time baselines found by the Southern radial velocity programmes of the Anglo-Australian Planet Search (e.g., Fig. 2 of \citealt{butler01}) and its continuous presence on the HARPS Guaranteed Time Observation list\footnote{e.g. \href{http://www.eso.org/sci/observing/teles-alloc/gto.html}{http://www.eso.org/sci/observing/teles-alloc/gto.html}}.

This paper is structured as follows. We introduce the RV data in section \ref{sec:data}. In section \ref{sec:model}, we introduce differential RVs to model wavelength-dependent noise and choose the so-called Goldilocks noise model in the Bayesian framework. In section \ref{sec:results}, we apply our models to the data and report the results. Finally, we discuss and conclude in section \ref{sec:conclusion}. 

\section{HARPS Doppler measurements for HD20794}\label{sec:data}
\begin{figure}
  \centering
  \includegraphics[scale=0.45]{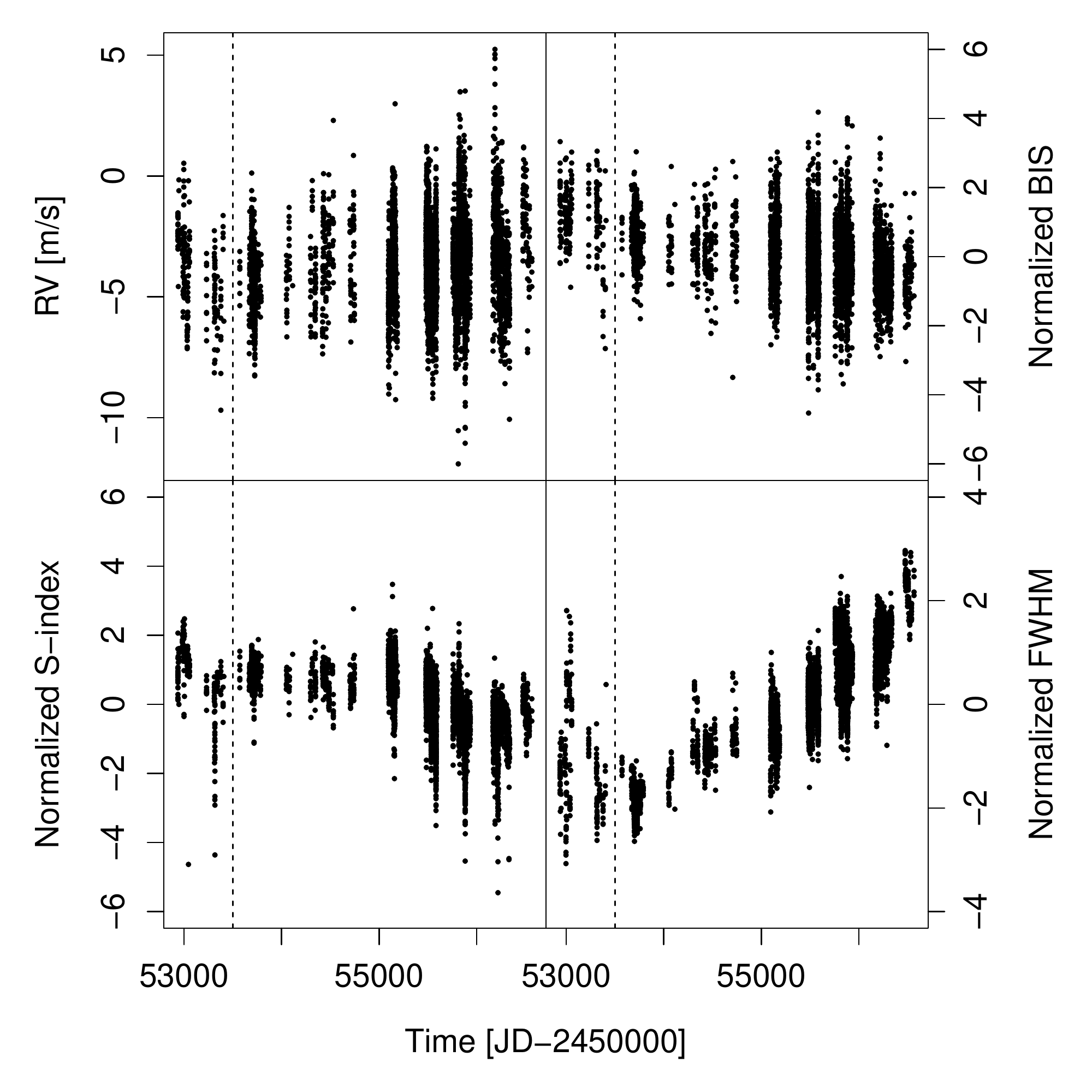}
  \caption{The measurements of RV, S-index, BIS and FWHM by HARPS for HD20794. The  dashed line denotes JD2453500 before which the data is probably affected by instrumental effects. }
  \label{fig:data}
\end{figure}

In the European Southern Observatory archive, there are 5150 publicly available RVs measured by HARPS from September 2003 to September 2013 for HD20794 from programmes Mayor, 60.A-9036, 072.C-0488, 083.C-1001, 084.C-0229, 086.C-0230, 087.C-0990, 088.C-0011, 089.C-0050, 090.C-0849, 091.C-0936. We process the Doppler measurements using the TERRA algorithm \citep{anglada12}, which generates RVs for 72 echelle orders. The stellar activity is recorded by the following indices: the S-index measured from the Ca II H\&K emission, line bisector shapes (BIS), and the width of the spectral lines (FWHM). We exclude the points with RVs or activity indices which deviate from the mean by more than 5-$\sigma$. We do this iteratively to omit any visible outliers. Moreover, we also remove the data obtained before JD2453500 due to the large dispersion in the activity indices, FWHM in particular. Such a feature is also observed in the $\tau$ Ceti data \citep{feng17a}, although the cause for this may be different. Finally, we obtain a data set of 4882 RVs together with activity indices. These data and the discarded epochs are shown in Fig. \ref{fig:data}. We see a great variation of FWHM before JD2453500, and this feature is also found in the data measured for $\tau$ Ceti \citep{feng17a}. The data is available in \url{http://star-www.herts.ac.uk/~ffeng/HD20794_supplementary/data}. 

\begin{figure*}
  \centering
  \includegraphics[scale=0.5,angle=-90]{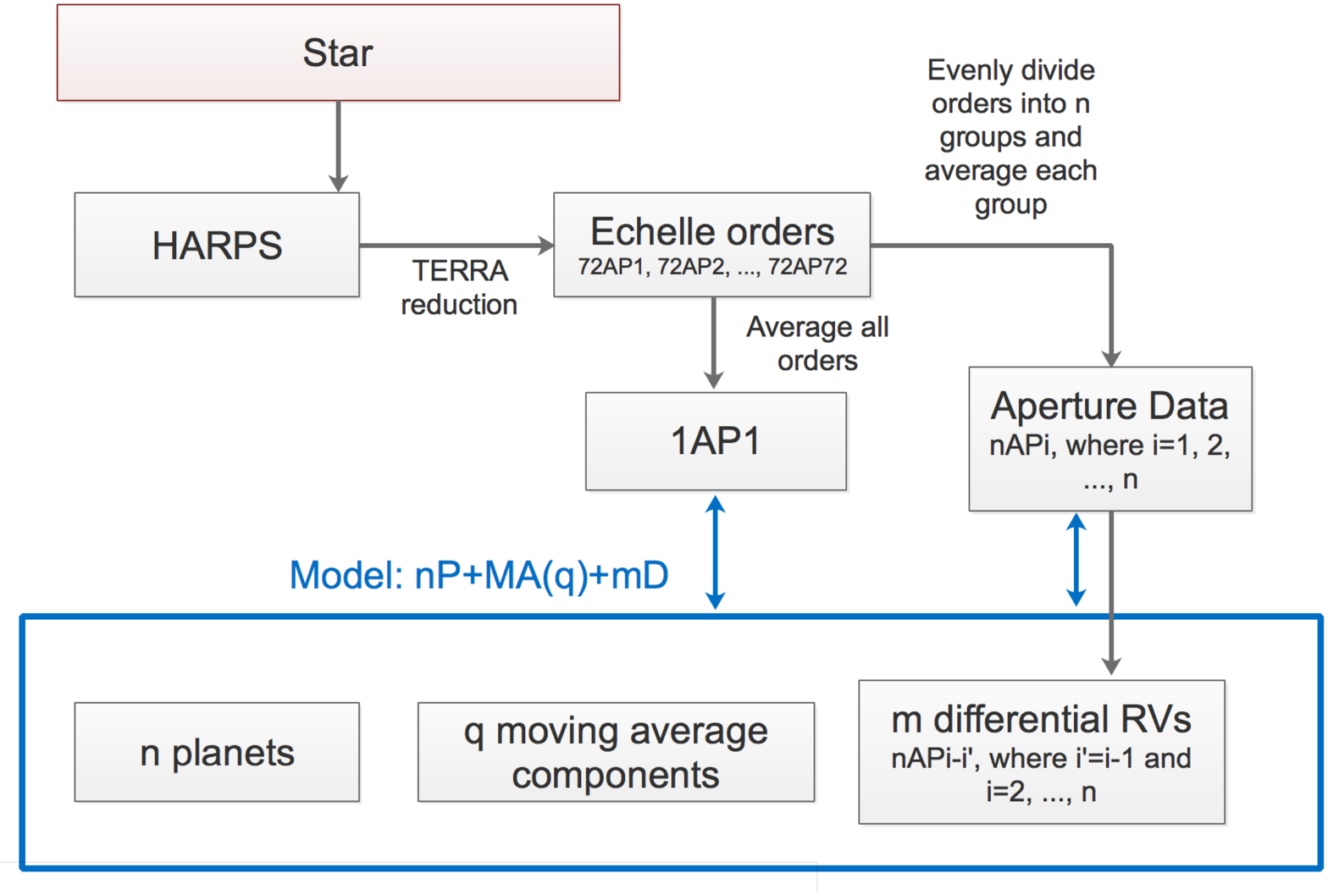}
  \caption{The data reduction and modeling process.}
  \label{fig:diagram}
\end{figure*}

We use the RV differences between the 72 echelle orders to remove wavelength-dependent noise. We divide the 72 spectral orders into groups, and average the RVs in each group to generate the so-called ``aperture data sets''. For $n$ groups of spectral orders, we generate $n$AP$j$ aperture data sets, where $j$ is a natural number not larger than $n$. We define the RV differences between aperture data sets as differential RVs. We denote them by $n$AP$j-i$, where $n$ is the number of groups of spectral orders, and $j$ and $i$ denote different aperture data sets. For $n$ independent aperture data sets, there are $n-1$ independent differential RVs in total. The schematic of this data reduction process is shown in Fig. \ref{fig:diagram}. 

\section{Noise model framework and Bayesian inference}\label{sec:model}
\subsection{Wavelength-dependent noise models}\label{sec:dRV}
The RV variation induced by stellar activity and rotation is typically wavelength-dependent \citep{tuomi13b}. Thus the average of RVs measured at different wavelength ranges (or spectral orders) is biased due to the ignorance of such a dependence. To remove such bias, we introduce ``differential RVs'' to weight the spectral orders {\it a posteriori}. Differential RVs are RV differences between different spectral orders (or wavelength ranges) and thus do not contain Keplerian variation since Keplerian variation is color-independent. Following the Goldilocks principle of noise modeling introduced by \cite{feng16}, we reduce the number of different RVs by deriving them from averaged spectral orders, so-called aperture data sets (see Fig. \ref{fig:diagram}). 

  Then the differential RVs are 
\begin{equation}
  D(t_i,\lambda_j)\equiv v(t_i,\lambda_{j+1})-v(t_i,\lambda_j)=\Psi(t_i,\lambda_{j+1})-\Psi(t_i,\lambda_j)~,
\end{equation}
where $v(t_i,\lambda_j)$ is the RV measured at time $t_i$ within a wavelength range centered at $\lambda_j$. We define the noise component of the RV model as 
\begin{equation}
\hat{\Psi}(t_i,\lambda_j) =
a(\lambda_j)\,t_i+b(\lambda_j)+\sum_{k}c_k(\lambda_j)I_k+\sum_{m=1}^{N_\lambda-1}d_mD(t_i,\lambda_m)~,
\label{eqn:psi}
\end{equation}
where $N_\lambda$ is the number of wavelength ranges or aperture divisions and thus $N_D\equiv N_\lambda-1$ is the number of independent differential RVs, $d_m$ characterizes the linear dependence of RV noise on the $m^{\rm th}$ differential RV, $a$ is the linear acceleration caused by stellar companions or activity cycles and $b$ is the reference velocity. The linear dependence of RVs on activity index $I_k$ is parameterized by constant $c_k$. We consider that $a$ and $b$ can be wavelength dependent, because besides fitting for trends due to companions, those parameters can also absorbs any drift due to long term activity not correlated with activity indices or long term instrumental noise variation. We linearly combine $\hat{\Psi}(t_i,\lambda_j)$ with $n$ Keplerian components to define the basic RV model, which is
\begin{eqnarray}
  \hat{v}_b(t_i, \lambda_j)&=&\sum_{k=1}^{n} f_k(t_i)+\hat{\Psi}( t_i,\lambda_j)~,\\\nonumber
  f_k(t_i)&=&K_k [\cos(\omega_k + \nu_k(t_i))+e_k\cos(\omega_k)]~,
\label{eqn:basic}
\end{eqnarray}
where $f_k(t_i)$ is the $k^{\mathrm{th}}$ Keplerian component, and $\hat{\Psi}$ is the noise component. In the Keplerian function, $K_k$, $\omega_k$, $\nu_k$, $e_k$  are the
amplitude, longitude of periastron, true anomaly and eccentricity for the
$k^{\mathrm{th}}$ Keplerian signal. 

Since the moving average (MA) model can efficient account for red noise in RVs and avoid false negatives \citep{feng16}, we model the red noise using a general MA model with exponential smoothing which is
\begin{equation}
  \hat{v}(t_i,\lambda_j)=\hat{v}_b(t_i,\lambda_j)+\sum_k w_k(\lambda_j) \exp[-|t_i-t_{i-k}|/\tau(\lambda_j)] \epsilon_{i-k},
\label{eqn:full}
\end{equation}
where $w_k$ and $\tau$ are the amplitude and time scale of the moving
average, and $\epsilon_{i-k}$ is $v(t_{i-k},\lambda_{j})-\hat{v}_b(t_{i-k}, \lambda_j)$. Hereafter, we define ``nP+MA(q)+mD'' as the $n$-planet model combined with $q^{\rm{th}}$-order moving average and $m$ differential RVs deriving from $m+1$ aperture data sets (see Fig. \ref{fig:diagram}). The white noise model is $\hat{\Psi}(t_i,\lambda_j)$, denoted by MA(0). 

We also account for a white excess noise with an amplitude of $s_J$ in the likelihood which is
\begin{equation}
\mathcal{L}_j\equiv P(v_j|\boldsymbol{\theta},M)=\prod_i\frac{1}{\sqrt{2\pi[\sigma_{i,j}^2+s_J(\lambda_j)^2]}}\exp\left[-\frac{(\hat{v} (t_i,\lambda_j)-v_{i,j})^2}{2(\sigma_{i,j}^2+s_J(\lambda_j)^2)}\right]~,
\label{eqn:like}
\end{equation}
where $\sigma_{i,j}$ is the measurement noise at time $t_i$ in the j$^{th}$ aperture data set with wavelength range centered at $\lambda_j$, $s_J(\lambda_j)$ is
the jitter level, and $v_{i,j}$ is the RV measured at time $t_i$ in the j$^{th}$ aperture data set. Following \cite{feng16}, we adopt a Gaussian prior distribution centered at zero and with a standard deviation of 0.1 for eccentricity, logarithmic uniform distributions for period $P$ and correlation time scale $\tau$, and uniform distributions for other parameters. The specific prior distributions are given in Table \ref{tab:prior}. 

\begin{table*}
  \centering
\caption{The prior distributions of model parameters. The 
  unit of $c_k$ and $d_m$ is m/s. The maximum and minimum time of the data are denoted by $t_{\textrm{max}}$ and $t_{\textrm{min}}$, respectively. The maximum RV with respect to the mean is denoted as $|v-\bar{v}|_{\rm{max}}$. The parameter characterizing the linear dependence of RV on activity indices is $c_k$, where $k$ denotes various indices.}
\label{tab:prior}
  \begin{tabular}  {c*{4}{c}}
\hline 
Parameter&Unit&Prior distribution& Minimum & Maximum\\\hline 
        \multicolumn{5}{c}{\it Each Keplerian signal}\\
$K_j$&m/s&$1/(K_{\text{max}}-K_{\text{min}})$&0&$2|v-\bar{v}|_{\text{max}}$\\
$P_j$&day&$P_j^{-1}/\log(P_{\text{max}}/P_{\text{min}})$&1&$t_{\text{max}}-t_{\text{min}}$\\
$e_j$&---&$\mathcal{N}(0,0.1)$&0&1\\
$\omega_j$&rad&$1/(2\pi)$&0&$2\pi$\\
$M_{0j}$&rad&$1/(2\pi)$&0&$2\pi$\\\hline 
        \multicolumn{5}{c}{\it Linear trend and jitter}\\
$a$&m\,s$^{-1}$yr$^{-1}$&$1/( a_{\text{max}}- a_{\text{min}})$&$-365.24K_{\text{max}}/P_{\text{max}}$&$365.24K_{\text{max}}/P_{\text{max}}$\\
    $b$&m/s&$1/( b_{\text{max}}- b_{\text{min}})$&$-K_{\text{max}}$&$K_{\text{max}}$\\
$s_J$ &m/s&$1/(s_{J\rm max}-s_{J\rm min})$&0&$K_{\rm max}$\\\hline 
\multicolumn{5}{c}{\it Moving average}\\ 
$w$&---&$1/(w_{\text{max}}-w_{\text{min}})$&-1&1\\
$\tau$&day&$P_j^{-1}/\log(\tau_{\text{max}}/\tau_{\rm{min}})$&$1/(t_{\rm{max}}-t_{\rm{min}})$&1\\\hline 
    \multicolumn{5}{c}{\it Activity indexes and differential RVs}\\
$c_k$&m/s&$1/(c_{k\text{max}}-c_{k\text{min}})$&$-c_{k\text{max}}$&$K_{\text{max}}/(I_{X\text{max}}-I_{X\text{min}})$\\
$d_m$\,($m\in\{1,...,N_\lambda-1\}$)&m/s&$1/(d_{m\rm{max}}-d_{m\text{min}})$&$-d_{m\rm{max}}$&$K_{\text{max}}/(D_{m\rm{max}}-D_{m\rm{min}})$\\\hline 
  \end{tabular}  
\end{table*}

In almost all previous research, only the averaged data or 1AP1 is analyzed without accounting for wavelength-dependent noise. We justify the usage of differential RVs in the removal of wavelength-dependent noise by applying the model defined in Eqn. \ref{eqn:full} and \ref{eqn:like} to the 3AP1, 3AP2, 3AP3 and 1AP1 data sets. We model these data sets using the noise models of 0P+MA(1)+0D, 0P+MA(1)+2D, 0P+MA(0)+0D, 0P+MA(0)+2D. We calculate the generalized periodogram with floating trend (GLST; \citealt{feng17b}) for each data set and each model to visualize the consistency of signals in wavelength. Following \cite{cumming99}, the GLST is normalized with respect to the residual variance and the FAP is calculated accordingly.

\begin{figure*}
  \centering
  \hspace*{-10mm}
\includegraphics[scale=0.3]{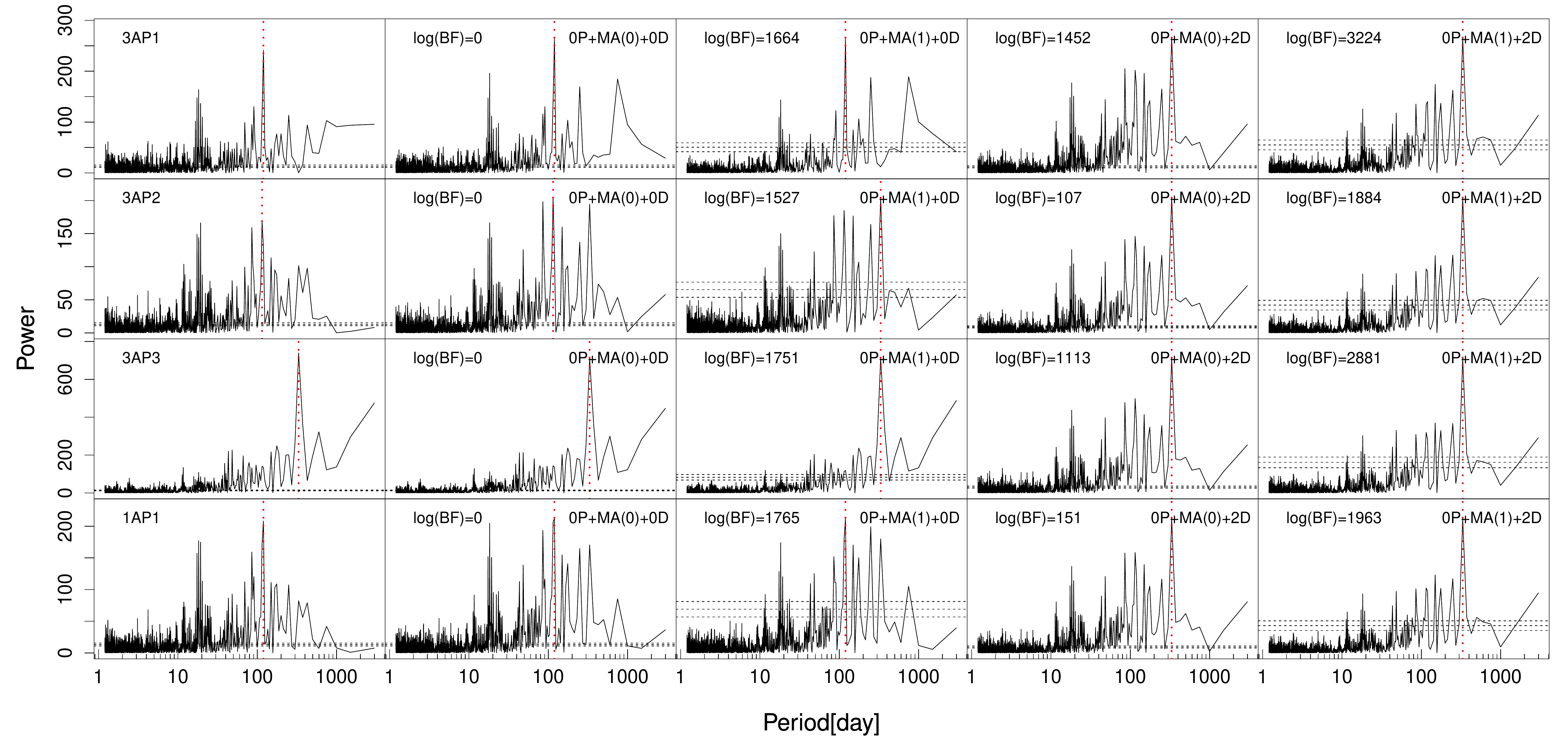}
  \caption{We aim to illustrate the existence of wavelength-dependent noise and the necessity of removing it using differential RVs. The interpretation of specific peaks in plotted GLSTs is not important. Rather, our concern is the consistency between GLSTs of RVs and their residuals measured at different wavelengths. Thus we show a series of GLSTs across the page. Each row of plots is given to a different aperture data set. Going down the page they are for the 3AP1, 3AP2, 3AP3 and 1AP1 data sets. The following columns show GLSTs for the corresponding residuals after subtracting the best-fit models of 0P+MA(0)+0D (white noise), 0P+MA(1)+0D (moving average), 0P+MA(0)+2D (white noise with differential RVs),  0P+MA(1)+2D (moving average with differential RVs). The logarithm BF of a model with respect to the MA(0) model are shown for the residuals of each data set after subtracting the best model prediction. In each panel, the red dotted line denotes the period with the highest power. The false alarm probability (FAP) threshold of 0.1, 0.01 and 0.001 are shown by dashed lines. The FAPs are calculated assuming the null hypothesis of white noise and thus are not used to measure the significance of signals in the data contaminated by correlated noise.}
\label{fig:residual}
\end{figure*}

The GLSTs for the 3AP1, 3AP2, 3AP3 and 1AP1 data sets and their residuals after subtracting the best-fit noise model are shown in Fig. \ref{fig:residual}. We observe significant differences between GLSTs for 3AP1, 3AP2, 3AP3 and 1AP1, indicating significant wavelength-dependent noise in the RV data. On the contrary, the residuals of all RV data sets have similar GLSTs after subtracting the best-fit 0P+MA(0)+2D or 0P+MA(1)+2D from the data. Such consistency is not sensitive to the change of time-correlated noise component (i.e. MA(0) and MA(1)), as seen in the rightmost two columns. Therefore differential RVs are essential in removing wavelength-dependent noise.

In the following sections, we will focus our analysis on the 1AP1 data set because it has a higher signal to noise ratio with respect to the other aperture data sets. As seen in Fig. \ref{fig:residual}, the wavelength-dependent noise in 1AP1 is properly removed by differential RVs {\it a posteriori}. 

\subsection{Goldilocks noise model}\label{sec:goldilocks}
We apply the adaptive Metropolis Markov Chain Monte Carlo (AM) introduced by \cite{haario01} to sample the posterior. Specifically, we use tempered (or hot) AM chains to identify potential signals, and then use un-tempered (or cold) chains to constrain signals. Similar methods have been introduced by \cite{gregory11}. The hot chain is defined as a chain moving according to the modified posterior $P({\mathcal D}|\theta,M)^{\beta}P(\theta|M)$ with $0<\beta\leq1$, where $P({\mathcal D}|\theta,M)$ and $P(\theta|M)$ are the likelihood and prior of parameters $\theta$ for model $M$ and data ${\mathcal D}$. We use four parallel chains for inference if the posterior samples drawn by these chains converge to a stationary distribution. In other words, the ratio of variances between and within these chains should be less than 1.1, which is the so-called ``Gelman-Rubin criteria'' \citep{gelman92}. We typically use the AM algorithm to draw a few million posterior samples for model and/or parameter inference. The detailed results are available online: \url{http://star-www.herts.ac.uk/~ffeng/HD20794_supplementary/results}.

From the likelihoods of the posterior samples drawn by cold chains, we estimate the Bayes factor (BF) using the Bayesian information criterion (BIC). Following \cite{feng16}, we regard model A as favored over model B if the logarithmic BF of the former with respect to the latter is larger than 5. To confirm a signal, we also adopt the criteria used by \cite{tuomi12b} that the period should be constrained from above and below in the posterior distribution of the period. In other words, $P(P_k|{\mathcal D},~M)$ converges to a stationary distribution. 
  
Following the Goldilocks principle introduced by \cite{feng16}, we apply the BF criterion to determine which noise model to use for removing noise. We generate a set of noise models by setting $q\in\{0,1,2,3,4,5,6\}$ and $N_{D}\in\{0,2,5,8\}$, and apply them to the data. The results of the model comparison are shown in Table \ref{tab:noise}. We find that the fourth order MA combined with five differential RVs is the best choice. Apart from $q$ and $N_D$, all other model parameters are free. Specifically there are 2 parameters for the trend, 1 for jitter, 3 for activity indices, 5 for differential RVs and 5 for the MA model. There are 16 free parameters in total in the 0P+MA(4)+5D model. In the following section, we analyze the HARPS data using this model. 
\begin{table*}
  \centering
\caption{The logarithmic BIC-estimated BFs for noise models with respect to the model without any MA components or differential RVs. For all models, the BIS, FWHM and S-index are linearly combined with differential RVs.}
\label{tab:noise}
\begin{tabular}{c|*{7}{c}}
  \hline
$N_D$&MA(0)&MA(1)&MA(2)&MA(3) &MA(4)&MA(5)&MA(6) \\\hline
0 & 0 &1769 &1969& 2006& 2011& 2008& 2008\\
2 &151 &1963 &2219 &2288 &2299 &2299 &2299\\
5 &238 &1985 &2241 &2312 &2325 &2324 &2323\\
8 &236 &1976 &2233 &2305 &2318 &2317 &2316\\\hline
 \end{tabular}
\end{table*}

\section{Keplerian candidates}\label{sec:results}
We apply both Keplerian and sinusoidal functions to model the RV variations caused by planets, which are called ``Keplerian solution'' and ``circular solution'', respectively. We identify six signals in both solutions, and report the BFs for them in Table \ref{tab:BF}. Although the $\sim$43\,d signal identified in the Keplerian solution does not pass the BF threshold of 150, it satisfies the criteria for the circular solution. Moreover, without including the 43\,d signal into the model, the eccentricity of the 89\,d signal would be as high as 0.4. The 43\,d and 89\,d signals are approximately in 1:2 resonance, which is not rare for exoplanets \citep{steffen14}. Although the signal at a period of 11\,d does not pass the BF threshold in both solutions, it is strong in the Keplerian solution, and increases the BF by at least one order of magnitude. In addition, it is visible in the periodograms for all observation seasons, as we will see later. The periods in the circular solution are different from the ones in the Keplerian solution because of the assumption of periods in the circular solution. 

\begin{table*}
   \centering
\caption{The logarithmic BIC-estimated BFs for models with various numbers of planets for Keplerian and circular solutions.}
\label{tab:BF}
\begin{tabular}{c|*{6}{c}}
  \hline
    Number of planets&1&2&3&4&5&6\\\hline
    Period in Keplerian solution&332.69&18.33 &88.98 &147.08&11.86&43.14\\
    $\log(\rm BF)$& 38&  72&  96& 134& 137&135\\\hline
    Period in circular solution&337.26&18.32&86.80&147.33&43.33&11.29\\
    $\log(\rm BF)$&46  &83  &95 &121 &145 &147\\\hline
 \end{tabular}
\end{table*}

\begin{figure*}
  \centering
\includegraphics[scale=0.5]{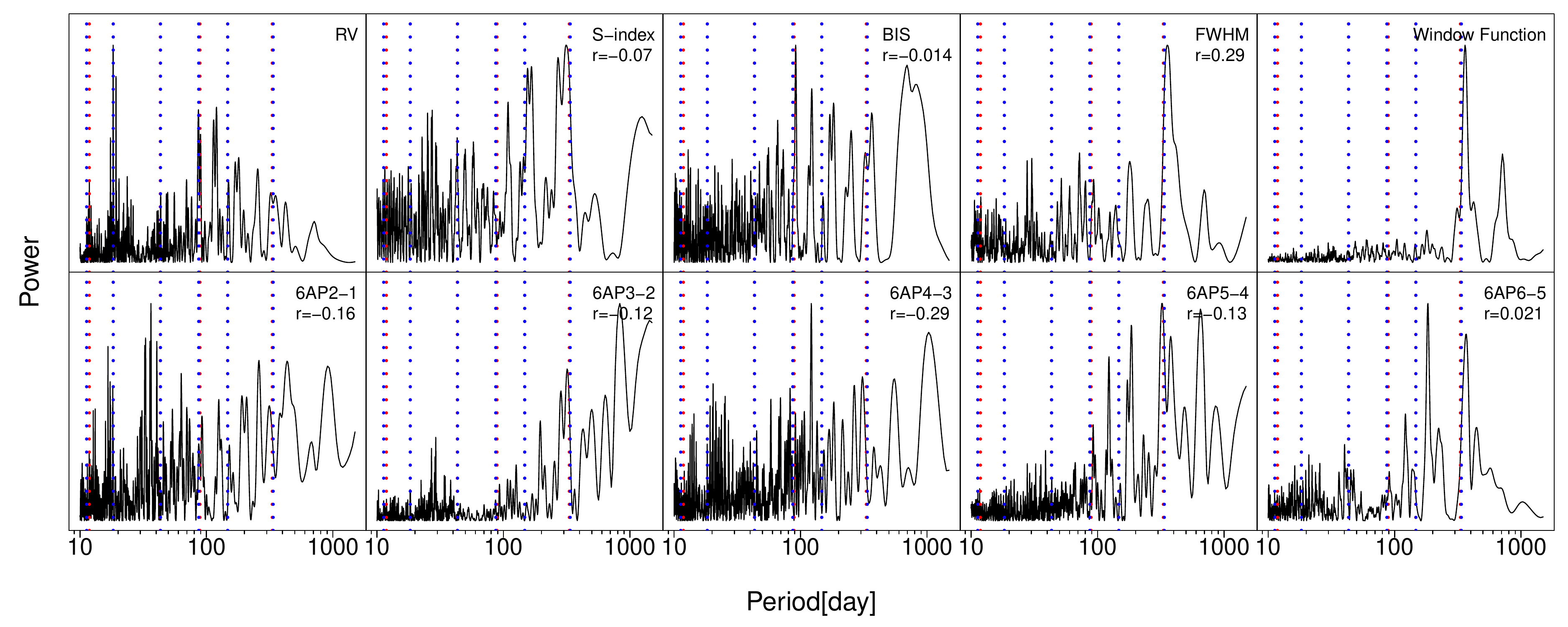}
  \caption{The GLSTs for the 1AP1 data set, various noise proxies and the window function. The names of the times series are shown at the top left corner. The red and blue dotted lines denote the signals quantified for the 6-planet model in the Keplerian and circular solutions respectively and reported in Table \ref{tab:BF}. The blue and red lines are usually too close to be distinguished for signals at periods larger than 12\,d. We truncate the period range for optimal visualization of signals. We did not show the FAPs because the highly correlated noise in the data is not accounted for by white noise FAPs.}
  \label{fig:indices}
\end{figure*}

To test whether the identified signals are caused by stellar and instrumental noise, we show the GLSTs for the data, the activity indices and differential RVs in Fig. \ref{fig:indices}. In the figure, we find that the RVs are strongly correlated with the FWHM and differential RVs, demonstrating the essential role of these noise proxies in removing correlated noise. We also find that periods around the $\sim$330\,d signal are strong in noise proxies such as FWHM and 6AP5-4, though these periods typically deviate from 330\,d by more than 20 d. Such approximate overlaps are expected since the 330\,d signal is close to the annual sampling frequency. Moreover, the probability of a random overlap between a signal and significant powers in the GLSTs of noise proxies increase with the number of proxies. Considering the minimization of correlated noise using noise proxies in the model, the 330\,d signal is too strong to be caused by pure noise.

\begin{figure}
  \centering
\hspace{-0.3in}\includegraphics[scale=0.48]{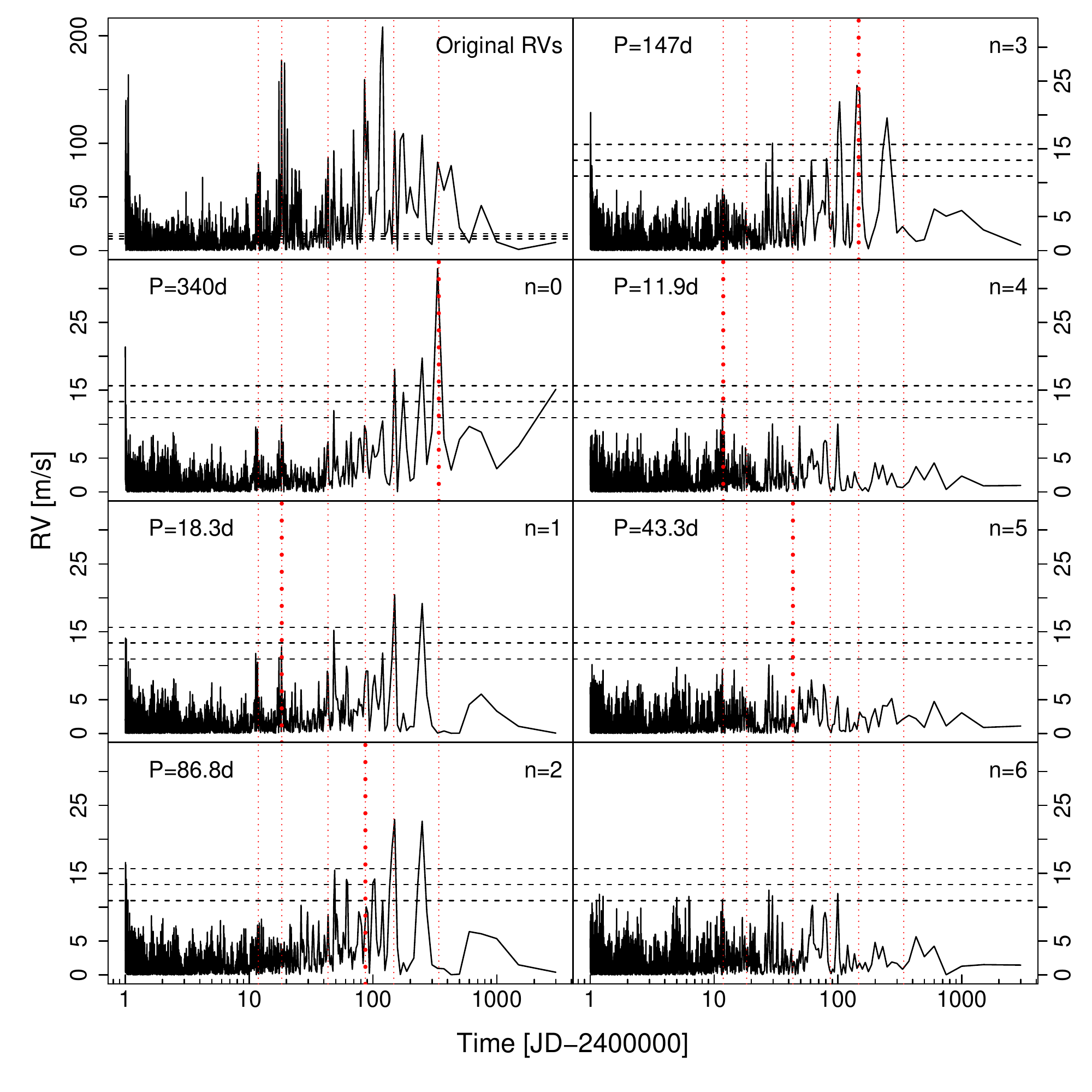}
  \caption{The GLSTs for the original data and data subtracted by the optimal prediction of $n$-planet model. The value of $n$ is shown in the upper right corner. The new signal identified by the $n+1$-planet model is shown by big red dotted lines. The other elements are similar to those in Fig. \ref{fig:indices}.}
  \label{fig:res_periodogram}
\end{figure}

We also calculate the GLSTs for the data subtracted by the optimal prediction of models with different numbers of Keplerian components, and show them in Fig. \ref{fig:res_periodogram}. We see that all identified signals correspond to certain peaks in the periodogram for the original RVs. The subtraction of the noise components makes some signals more significant. We find that the signal identified by posterior sampling is not always the most significant one shown in the residual periodogram. Moreover, the 43 and 11\,d signals are not strong in the corresponding residual periodograms, indicating the limitation of residual periodograms in identifying Keplerian signals. The residuals after subtracting the predictions of n-planet model are shown in Fig. \ref{fig:signal_residual} in the appendix. The correlation between the RV residuals and noise proxies are shown in Fig. \ref{fig:res_proxy}.

\begin{figure*}
\centering
\includegraphics[scale=0.6]{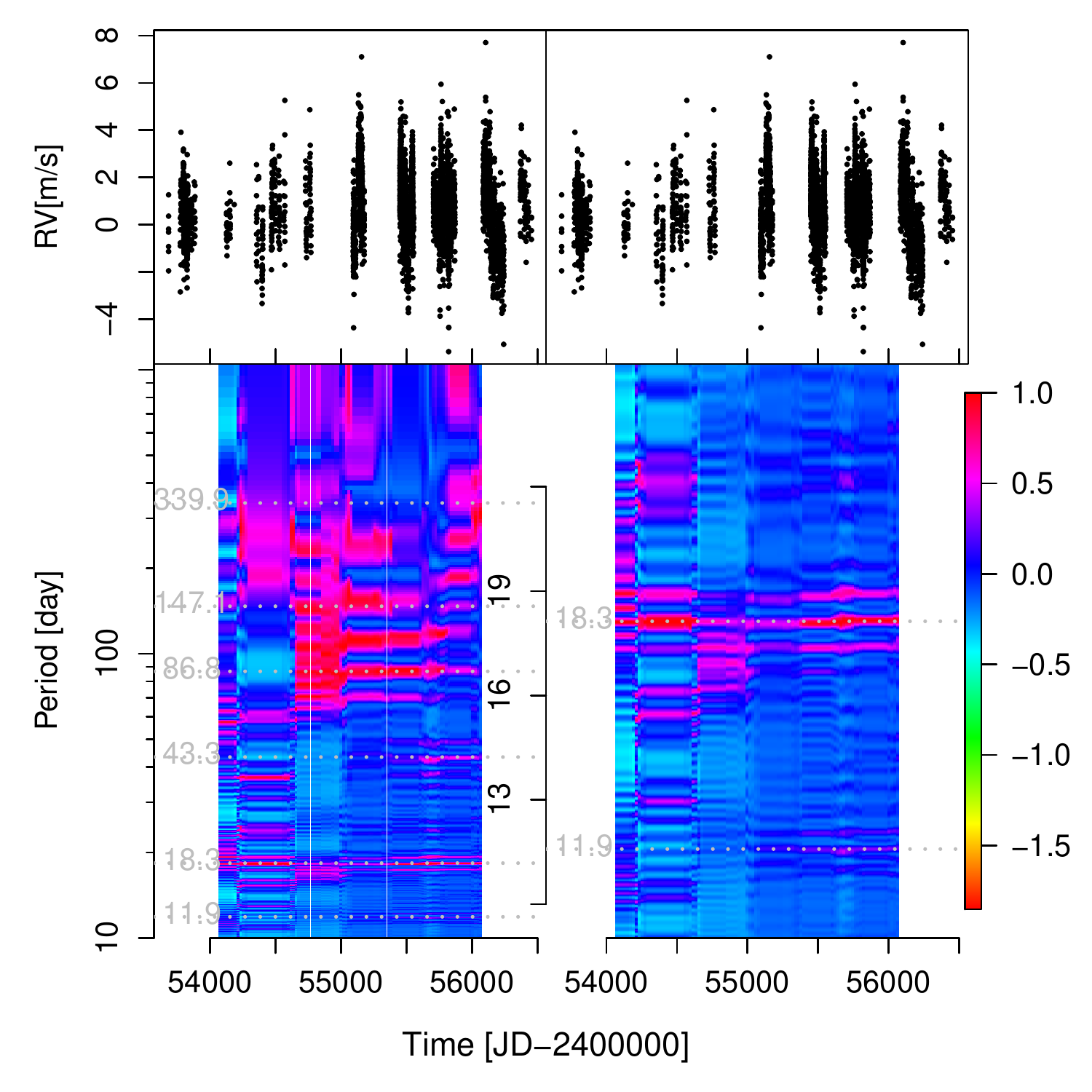}
\includegraphics[scale=0.6]{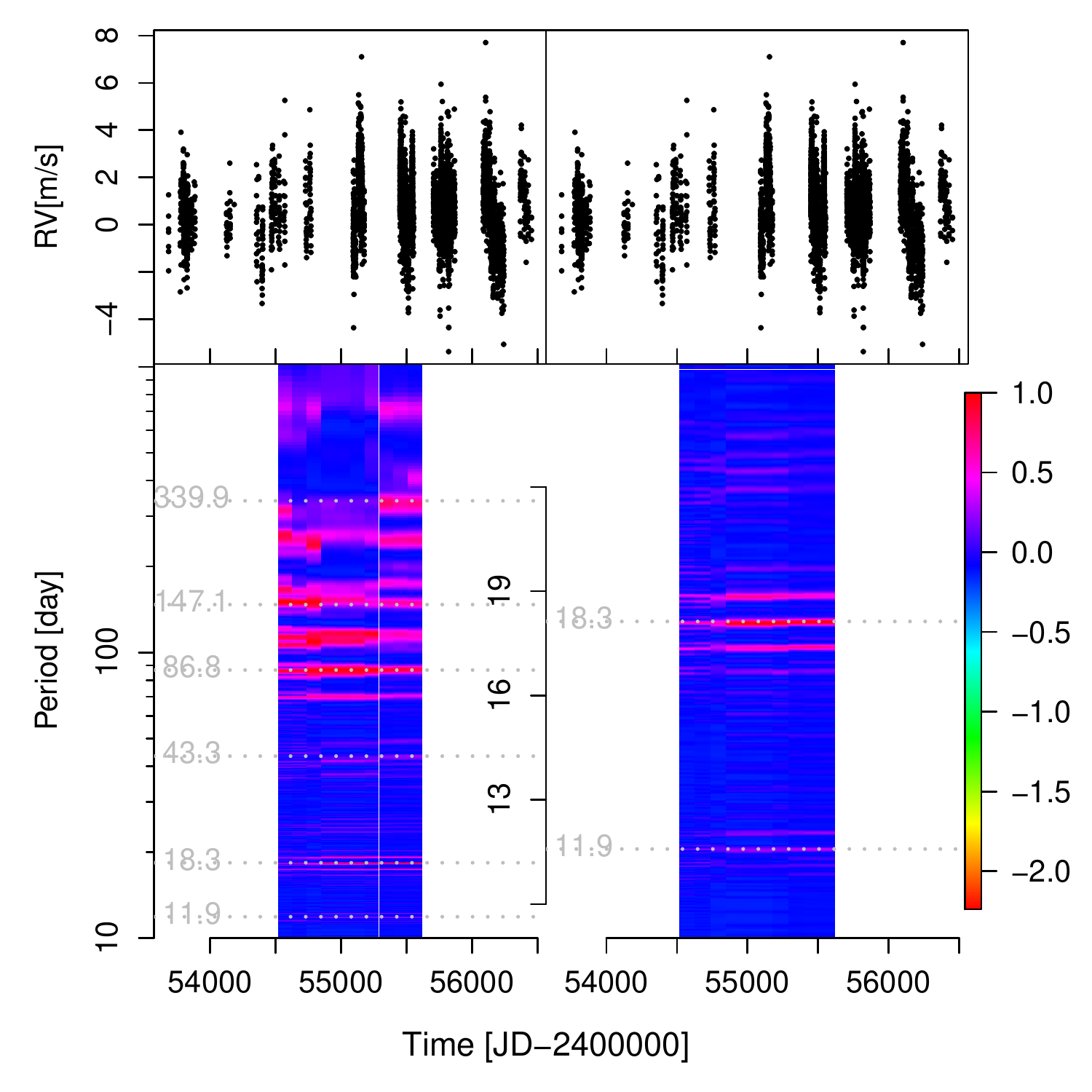}
\caption{The noise-subtracted RVs (top panels) and the moving MLP-based periodograms (bottom panels) made within the 1000\,d (left panels) and 2000\,d (right panels) moving window. In each moving periodogram, the bottom left panel shows the periodogram for periods ranging from 10\,d to 1000\,d while the bottom right panel is focused on visualization of periods below 30\,d. The colors encode the RML which is truncated to optimize the visualization of signals. The periods of signals identified in the 6-planet Keplerian solution are shown by horizontal dashed lines. The logarithmic ML is truncated to optimize the visualization of signals.}
\label{fig:periodogram}
\end{figure*}

To show the consistency of the detected signals, we make a Bayesian periodogram for data within a moving time window. To make this periodogram, we model the RVs using a combination of sinusoidal functions and a linear trend, and analytically marginalize the likelihood over the amplitudes of sinusoidal functions and the trend parameters which are $a$ and $b$ in Eqn. \ref{eqn:psi}. Hence this new periodogram is called ``marginalized likelihood periodogram'' (MLP; \citealt{feng17b}). In addition to trend parameter $a$, the MLP also optimize $b$, and thus extend the Bayesian generalized Lomb-Scargle periodogram \citep{mortier15}\footnote{The relevant code for MLP and GLST is available in Github: \url{https://github.com/phillippro/agatha}, based on which a web app is also developed and linked at \url{http://www.agatha.herts.ac.uk}}. To make a 2D periodogram, we move the time window in 100 steps to cover the whole data set, and compute the MLP for each step to construct a moving periodogram. We apply this method to the RVs from which the noise component in the 6-planet model prediction for the Keplerian solution has been subtracted. The moving periodograms with 1000\,d and 2000\,d time windows together with the data are shown in Fig. \ref{fig:periodogram}. Since more data is needed to cover the phase of long period signals, we recommend the 2000\,d time window for testing the consistency of the 330\,d and 147\,d signals. We calculate the MLP for each time step, and scale the logarithmic marginalized likelihood (ML) to be ${\rm RML}\equiv ({\rm ML}-\overline{\rm ML})/({\rm ML}_{\rm max}-\overline{\rm ML})$ where $\overline{\rm ML}$ and ${\rm ML}_{\rm max}$ are the mean and maximum ML. The RMLs are encoded by colors to show the consistency of signals in time rather than to estimate the significance of signals.

As we can see, the 11\,d signal together with its aliases, though weak, are visible over the whole time span. The 147 and 330\,d signals are consistently identified in the 2000\,d moving periodogram. Notably the significance of these signals may vary with time because long period signals are sensitive to data sampling and size, which typically vary with time. There are also strong powers around the harmonics of the 330\,d signal at a period of about 680\,d and the annual alias of the 147\,d signal at a period of about 250\,d. Considering that the 330\,d signal is also strong in some noise proxies, further observations and analyses are required to explore the nature of this long period signal. 

The 147\,d signal is strong in many observation seasons despite the slight variation of its period. The high periodogram power around 110\,d is probably a combination of the aliases of 89 and 147\,d since the subtraction of the 89\,d signal would reduce its significance. The signals around 89\,d, 43\,d and 18\,d are also consistently identified over the whole time span. These signals are previously found by P11 in the circular solutions. Notably the periods of $\sim$43 and 147\,d slightly deviate from the optimized values shown in the periodogram because the periodogram assumes zero eccentricity. In addition to the identified signals, the periodogram maps in Fig. \ref{fig:periodogram} also show annual alias of 330\,d around 174\,d, and annual aliases of 89\,d around 113 and 70\,d. The annual aliases of 43, 18 and 11\,d are also visible. Therefore we interpret all strong signals based on the moving periodograms which visualize the genuine signals as well as their aliases. 

\begin{figure*}
\centering
\includegraphics[scale=0.4]{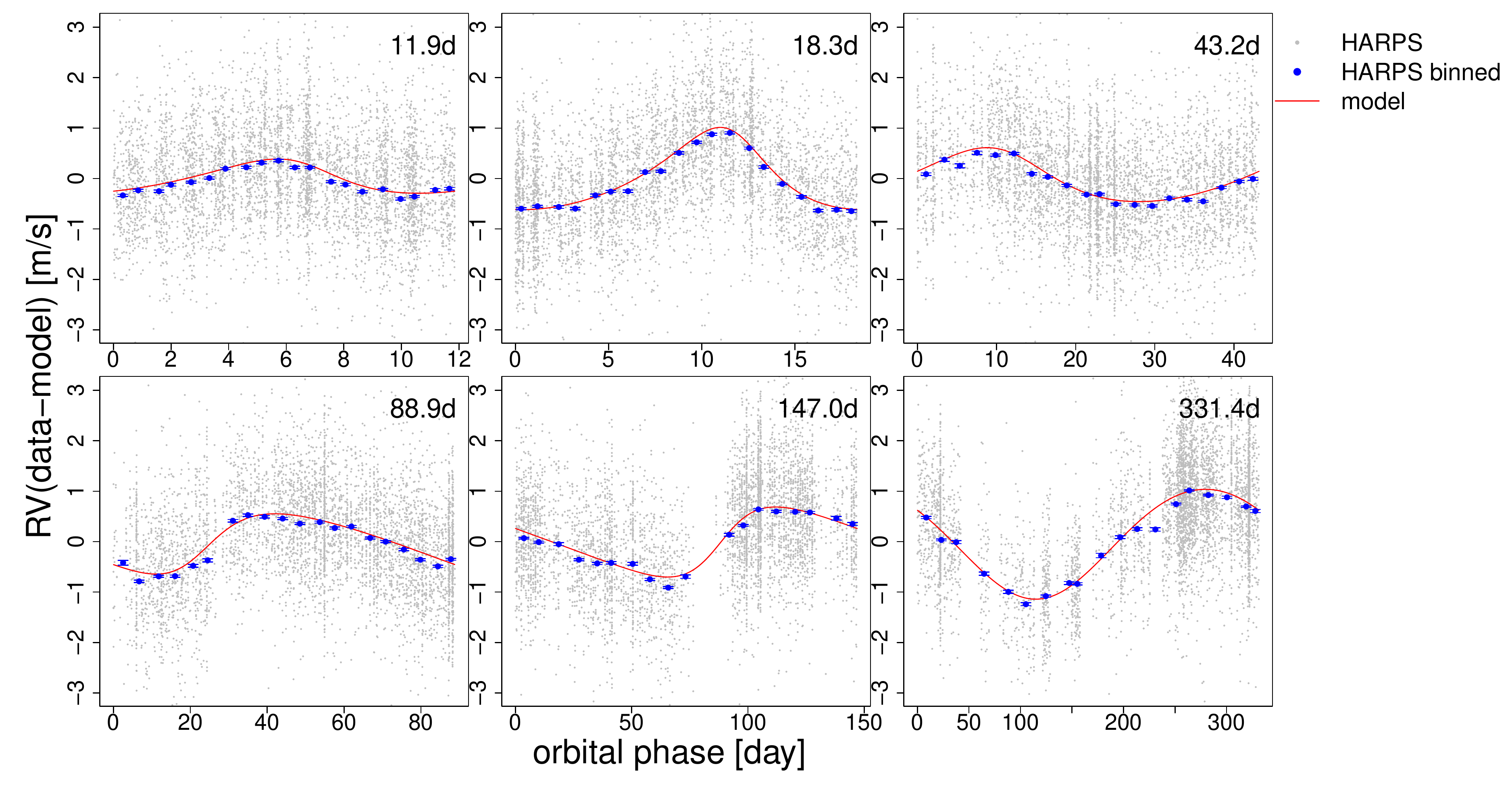}
\caption{The phase-folded data and the 6-planet model prediction. For the signal shown in each panel, the noise components and other signals are subtracted from the data. For each panel, the phase-folded times are original observation times mod the optimal period with respect to the minimum observation time.}
\label{fig:phase_curve}
\end{figure*}

Based on the above analyses, we interpret the signals around 18, 89, 147 and 330\,d as planet candidates, though investigations of the 330\,d signal is necessary to fully confirm its Keplerian nature. Since there is only weak evidence for the signals at periods of 11 and 43\,d, more data and further analyses are required to confirm these potential candidates. We show the phase curves for these six signals in Fig. \ref{fig:phase_curve}. We see that the 6-planet model well fits the data, supporting the existence of 6 signals despite considerable eccentricity for signals around 18, 43 and 147\,d. 

In Table \ref{tab:signal}, we report the parameters of the noise model and the four planet candidates, which are estimated in the 6-planet Keplerian solution. In the table, we see high MAP values of linear parameters such as $c_3$ and $d_3$\footnote{These parameters measure the contribution of noise RV variations and are in units of m/s because the proxies are normalized. On the other hand, the Pearson coefficients shown in Fig. \ref{fig:res_proxy} measure the correlation between RVs and proxies.}, suggesting the existence of correlated noise as high as 1\,m/s. This correlated noise would prevent any reliable detections of sub-meter signals if not minimized through the application of noise proxies such as activity indices and differential RVs. However, the application of noise proxies may introduce extra noise in the fitting \citep{feng17a} due to an oversimplification of the complex/nonlinear dependence of RVs on proxies by linear functions. This is also part of the reason why we do not interpret the weak signals at periods of 11 and 43\,d as planet candidates. On the other hand, the MA model plays an important role in removing time-correlated noise, which is probably intra-night RV variability caused by instrumental or reduction-process effect \citep{berdinas16}. 
\begin{table*}
  \centering
  \caption{The maximum {\it a posteriori} (MAP) estimation of the parameters for the noise model (lower rows) and the five signals (upper rows) detected in the TERRA-reduced HARPS data for HD20794. The parameters are estimated in the 6-planet Keplerian solution. The uncertainties of parameters are represented by the values determined at 1\% and 99\% of the cumulative posterior density. We estimate the minimum planetary mass and semi-major axis using a stellar mass of 0.813\,$^{+0.018}_{-0.012}M_\odot$ \citep{ramirez13}. Among the parameters of the noise model, $c_1$, $c_2$ and $c_3$ are linear coefficients for S-index, BIS and FWHM, respectively. }
\label{tab:signal}
\tiny
\hspace*{-8mm}
 \begin{tabular}  {c*{6}{c}}
\hline 
Parameters&       HD20794 b&      HD20794 d &     HD20794 e& HD20794 f?&  HD20794 g?&  HD20794 c?\\\hline
$P$\,(d)&18.33 [18.31, 18.34]&88.90 [88.49, 89.27]&147.02 [146.11, 148.45]&331.41 [328.40, 336.49]&11.86 [11.84, 11.87]&43.17 [43.07, 43.29]\\
    $K$\,(m/s)&0.81 [0.57, 0.81]&0.60 [0.42, 0.70]&0.69 [0.56, 0.84]&1.09 [0.94, 1.28]&0.34 [0.24, 0.50]&0.53 [0.36, 0.64]\\
$e$&0.27 [0.05, 0.31]&0.25 [0.04, 0.41]&0.29 [0.11, 0.43]&0.05 [0.00, 0.11]&0.20 [0.01, 0.35]&0.17 [0.01, 0.27]\\
$\omega$\,(rad)&6.74 [5.75, 7.22]&4.41 [3.64, 5.10]&4.68 [4.11, 5.52]&3.24 [0.91, 5.64]&0.77 [-1.47, 2.05]&0.55 [-1.60, 1.91]\\
$M_0$\,(rad)&2.26 [1.56, 3.07]&4.69 [3.89, 5.60]&2.52 [1.76, 3.46]&4.02 [1.00, 5.72]&2.74 [0.96, 4.74]&-1.67 [-2.87, 1.68]\\
$m\sin{i}$\,($M_\oplus$)&2.82 [2.02, 2.92]&3.52 [2.51, 4.10]&4.77 [3.91, 5.73]&10.26 [8.79, 12.15]&1.03 [0.73, 1.52]&2.52 [1.69, 3.04]\\
    $a$\,(au)&0.127 [0.126, 0.128]&0.364 [0.360, 0.368]&0.509 [0.503, 0.515]&0.875 [0.865, 0.886]&0.095 [0.094, 0.096]&0.225 [0.222, 0.227]\\\hline\hline
    $a$ (m\,s$^{-1}$\,yr$^{-1}$)&$b$ (m\,s$^{-1}$)&$c_1$ (m\,s$^{-1}$)&$c_2$ (m\,s$^{-1}$)&$c_3$ (m\,s$^{-1}$)&$s_J$ (m\,s$^{-1}$)&$w_1$\\
-0.52 [-0.64, -0.47]&-1.69 [-2.02, -1.12]&-0.07 [-0.13, 0.01]&0.02 [-0.01, 0.06]&0.93 [0.82, 1.06]&0.89 [0.87, 0.92]&0.41 [0.37, 0.44]    \\\hline
$w_2$&$w_3$&$w_4$& $\ln(\tau)$ (day)& $d_1$ (m\,s$^{-1}$)& $d_2$ (m\,s$^{-1}$)&$d_3$ (m\,s$^{-1}$)\\
0.23 [0.20, 0.28]&0.19 [0.14, 0.23]&0.14 [0.07, 0.16]&-2.06 [-2.37, -1.75]&-0.15 [-0.20, -0.13]&-0.35 [-0.40, -0.31]&-0.40 [-0.46, -0.36]  \\\hline
$d_4$ (m\,s$^{-1}$)&$d_5$ (m\,s$^{-1}$) &&&&&\\
-0.36 [-0.42, -0.31]&-0.23 [-0.28, -0.19]&&&&&\\\hline
   \end{tabular}
\end{table*}

In Table \ref{tab:signal}, the signal at a period of 147\,d corresponds to a super-Earth perturbing HD20794 with a mean semi-amplitude\footnote{The mean semi-amplitude is the semi-amplitude $K$ multiplied by $1-e$. } of about 0.5\,m/s. The candidate with 89\,d period has a minimum mass lower than the candidate around 90\,d reported by P11 probably because we have accounted for correlated noise which might have been interpreted as signals by P11. The 11\,d signal caused a radial velocity variation of about 0.35\,m/s. Regardless of whether this signal is Keplerian or not, the detection of such a weak signal demonstrates the important role of noise modeling in detecting low mass planets using the radial velocity technique.

According to the results of RV challenge, the MA model in combination of Bayesian methods can identify signals with a $K/N$ ratio as low as 5 without announcing a false positive \citep{dumusque16b}. The K/N ratio for signal with semi-amplitude of $K$ is defined as $K/N\equiv K/RV_{\rm rms}\times \sqrt{N_{\rm obs}}$, where $RV_{\rm rms}$ is the standard deviation of RVs after removing the best-fit trend and correlation with noise proxies, and $N_{\rm obs}$ is the number of observations. If we consider measurements in a 15\,min bin as an independent observation \citep{mayor03,otoole07}\footnote{The RVs were measured with high cadence and are thus highly correlated in time due to stellar oscillations and systematic errors \citep{teixeira09}. It is necessary to bin the data to calculate the number of independent observations. }, we get 713 independent observations and $RV_{\rm rms}=1.61$\,m/s. Then the $K/N$ ratios are 12, 10, 10, 19, 6.7 and 9.6 for signals at periods of 18, 89, 147, 330, 11 and 43\,d. Hence the sub-meter signals we have identified for HD20794 b, d, and e are well above the $K/N=7.5$ threshold of reliable detections of RV signals. Even the smallest signal at 11\,d is close to this threshold and is higher than the $K/N=5$ limit reached by our team. 

However, the nominal eccentricities of HD20794 b, d, and e are different from zero, probably leading to dynamical instability of the system according to the Lagrange stability analyses \citep{barnes06}. But we find that the system is stable if lower eccentricity within the uncertainty interval is adopted for HD20794 b, d and e. The high eccentricity reported in Table \ref{tab:signal} might arise from instrumental noise \citep{feng17a}. Specifically, the instrumental noise may cause short-term RV variations which favor high eccentricity solutions. The high eccentricity may also be caused by the intrinsic bias in estimating non-negative eccentricity \citep{zakamska11}. Therefore the proposed candidates probably have orbits more circular than those reported in this work.

Only the candidate with a period of about 330\,d is located in the habitable zone \citep{kopparapu13}, although the confirmation of this candidate requires further observations and analyses. Its habitability would be influenced by the frequent impacts of objects from the massive circumstellar disc \citep{kennedy15}.  All the other candidates are located too close to the host star to allow the existence of liquid water, and thus cannot be considered habitable. 

\section{Discussions and conclusions}\label{sec:conclusion}
We analyze the HARPS data of HD20794 in the Bayesian framework. We find strong dependence of the RV noise on wavelengths. This wavelength-dependent noise cannot be removed by averaging all spectral orders as previous studies did. To deal with this noise, we use differential RVs to weight the spectral orders {\it a posteriori}. We apply this method to data sets measured within different wavelength ranges, and find that differential RVs efficiently remove wavelength-dependent noise. Therefore we propose a combination of the MA model and differential RVs to remove the time and wavelength-dependent noise in RV datasets. 

By modeling the RV noise correlated in time and wavelength for HD20794, we identify three firm Keplerian candidates at periods of about 18, 89 and 147\,d. The signal at a period of 330\,d is probably Keplerian, although similar periods have strong powers in some proxies. There is also weak evidence for the existence of signals at periods of 43 and 11\,d. While the 43\,d signal does not pass the Bayes factor criterion, the inclusion of it in the model can reduce the eccentricity of the 89\,d signal. Although the 11\,d signal only increases the Bayes factor by a factor of about 20, it consistently appears in different data chunks. 

Thus by applying noise modeling to the P11 and newer HARPS data, we confirm the two candidates at periods of 18 and 90\,d reported by P11 and quantify them better. But we are suspicious about the existence of the reported 40\,d candidate because it does not pass the signal detection criteria. We also find a new planet candidates, HD20794 e, at a period of about 147\,d, corresponding to a super-Earth. The signal at a period of about 330\,d probably correspond to a Neptune located in the habitable zone of HD20794. The estimated orbital eccentricity of HD20794 b, d and e is larger than 0.2.

This considerable eccentricity is probably caused by instrumental noise and the fitting bias, as concluded by \citep{feng17a} and \cite{zakamska11}. Notably such eccentric solutions are also found in the HARPS data of $\tau$ Ceti \citep{feng17a}. This indicates a connection between high eccentricity and significant red noise in large RV datasets of bright stars that are observed with high cadence. Thus it is a viable hypothesis that the four planet candidates have eccentricities low enough to form a dynamically stable system. 

Our detection of signals with semi-amplitudes to below 0.5 m/s demonstrates the ability of the RV method to find Earth analogs orbiting Sun-like stars. This work supports the conclusion of \cite{dumusque16b} that the combination of Bayesian methods with the modeling of correlated noise is essential in finding small planets. In particular, the noise model framework developed by \cite{tuomi12} and \cite{feng17a} and applied in this work plays a key role in disentangling small signals from noise. 

\section*{Acknowledgements}
The authors are supported by the Leverhulme Trust (RPG-2014-281) and the Science and Technology Facilities Council (ST/M001008/1). We used the ESO Science Archive Facility to collect radial velocity data sets.  The authors gratefully acknowledge the HARPS-ESO team’s continuous improvement of the instrument, data and data reduction pipelines that made this work possible. Finally, the authors thank the referee, Xavier Dumusque, for his valuable comments that helped improving the manuscript significantly.

\appendix
\section{Residual visualization}

\begin{figure*}
  \centering
  \includegraphics[scale=0.8]{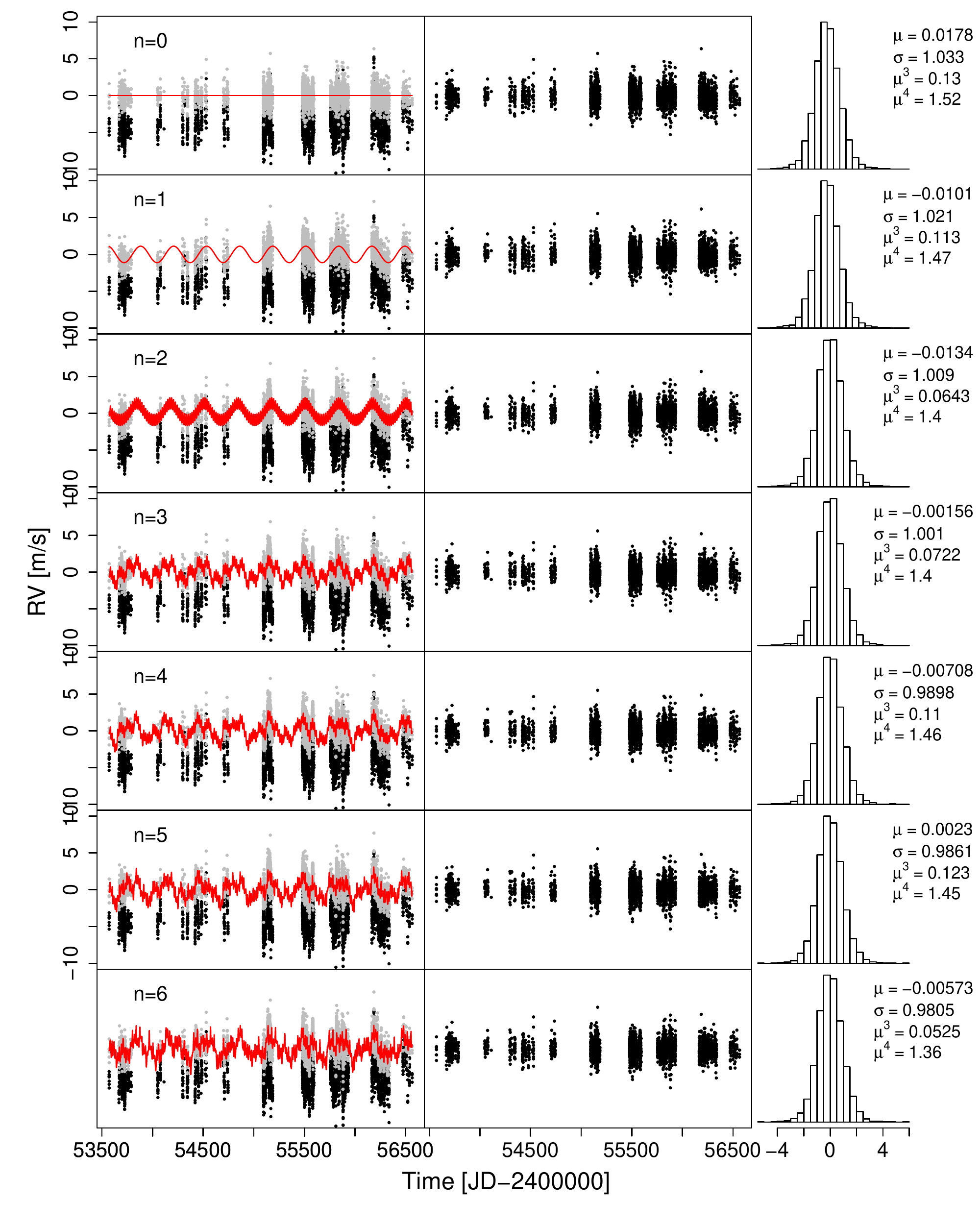}
  \caption{The signals and residuals for n-planet model. In the left panels, the original data, the noise-subtracted data and the signals are shown by black points, grey points and red curve, respectively. The residuals after subtracting signals and the noise model predictions are shown in the middle panels with the same RV range as in the left panels. The distributions of residuals are shown in the right panels. The mean ($\mu$), standard deviation ($\sigma$), skewness ($\mu^3$), and kurtosis ($\mu^4$) are shown for each distribution.}
  \label{fig:signal_residual}
\end{figure*}

\begin{figure*}
  \centering
  \includegraphics[scale=0.45]{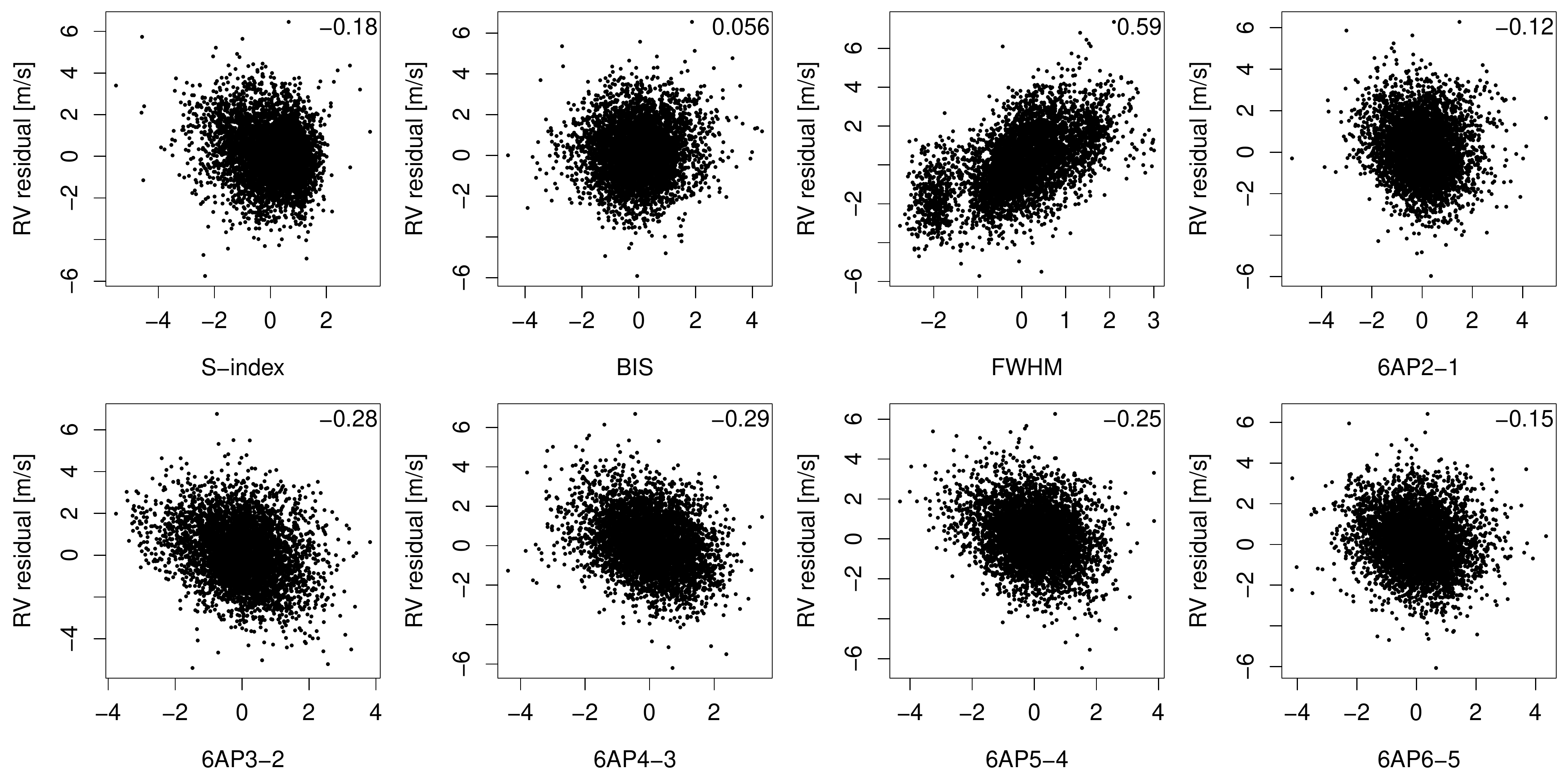}
  \caption{The correlation between RV residuals and noise proxies. The RV residuals for a given proxy are calculated by subtracting from the 1AP1 data set the best-fitting 6-planet model, where the linear parameter corresponding to the proxy is set to zero. The Pearson correlation coefficient is shown in the top right corner. }
  \label{fig:res_proxy}
\end{figure*}

\bibliographystyle{aa}
\bibliography{nm}
\end{document}